\documentclass[eprint,amsmath,amssymb,aps,nofootinbib]{revtex4-1}

\newcommand\simlt{\lower.5ex\hbox{$\; \buildrel < \over \sim \;$}}
\newcommand\simgt{\lower.5ex\hbox{$\; \buildrel > \over \sim \;$}}

\usepackage{graphicx}
\usepackage{dcolumn}
\usepackage{bm}

\begin{document}

\title{Loaded magnetohydrodynamic flows in Kerr spacetime}
\author{Noemie Globus}
\author{Amir Levinson}
\affiliation{School of Physics \& Astronomy, Tel Aviv University, Tel Aviv 69978, Israel}

\date{October 6, 2013}

\begin{abstract}
The effect of mass and energy loading on the efficiency at which energy can be extracted magnetically from a Kerr black hole is explored,
using a semi-analytic, ideal MHD  model  that incorporates plasma injection on magnetic field lines.   We find a critical load below which 
the specific energy of the plasma inflowing into the black hole is negative, and above which it is positive, and identify two types of flows with 
distinct properties;  at sub-critical loads a magnetic outflow is launched from the ergosphere, owing to extraction of the black hole spin energy, as originally proposed by Blandford and Znajek.  At super-critical loads the structure of the flow depends on the details of the injection process. 
In cases where the injected plasma is relativistically hot, a pressure driven, double trans-magnetosonic flow is launched from a stagnation point 
located outside the ergosphere, between the inner and outer light cylinders.  Some fraction of the energy deposited in the magnetosphere is then
absorbed by the black hole and the rest emerges at infinity in the form of a relativistic outflow.  When the injected plasma is cold an outflow may not form at all. We discuss the implications of our results to gamma ray bursts and active galactic nuclei. 
\end{abstract}


\maketitle

 

\section{Introduction}

A plausible production mechanism for the relativistic outflows observed in AGNs, GRBs, and microquasars is magnetic extraction of the spin energy of a Kerr black hole.  
It has been shown \cite{BZ77} that in the force-free limit, at which the inertia of the plasma is negligible, frame dragging induces an outward flow of energy  along magnetic field lines threading the horizon, at the expense of the black hole's rotational energy.  
It is commonly thought that this outward energy flux ultimately transforms into a collimated relativistic jet, like those seen in the compact relativistic systems mentioned above.  Indeed, recent numerical simulations (e.g., \cite{Kom04,Hretal04,McK04,McK05,BK08,KB09}) indicate that powerful outflows can be produced by this mechanism if sufficiently large magnetic flux  can be accumulated near  the horizon of the black hole. 

A question of interest is how the inertia of the plasma injected on magnetic field lines affects the properties of the emerging outflow, and in particular what are the requirements  for the activation of the Blandford Znajek  mechanism (hereafter BZ).  Takahashi et al. \cite{TNTT90} considered the structure of a  cold MHD inflow in Kerr spacetime, and have shown that two conditions must be fulfilled in order for energy to be extracted: (i) the angular velocity of magnetic field lines must satisfy $0<\Omega_F<\Omega_H$, where $\Omega_H$ is the angular velocity of the black hole, and (ii) the Alfv\'en point must be located inside the ergosphere.   Condition (ii) is automatically satisfied in the force-free limit, but not necessarily in general.  The question of how the location of the Alfv\'en point depends on the load was not addressed in \cite{TNTT90}. 

In this paper we show that there is a critical energy load below which the outflow is powered by the black hole, and above which it is either powered by the external energy source or does not form at all.   This critical load depends on the strength of magnetic field lines threading the horizon and the angular momentum of the black hole.  One immediate consequence is that the mass inflow that supports the magnetic field near the horizon must be strongly suppressed in the polar region in order for a BZ outflow to be launched.   A similar conclusion was drawn by Komissarov \& Barkov \cite{KB09}, who conducted numerical experiments to study the effect of mass loading on the energy extraction process in GRBs. They have shown that in the collapsar model the requirement  for the activation of the BZ process imposes stringent constraints on the progenitor star.  But even if the progenitor accommodates those requirements and the polar region is devoid of baryons, 
 substantial loading is anticipated owing to deposition of hot plasma by annihilation of neutrinos emanating from the accretion flow surrounding the black hole \cite{Jar96,PW99,BAJM07,zB11}.  Below we show that if the inward enthalpy flux of the hot plasma deposited in the magnetosphere exceeds a certain value, the
BZ process completely shuts down, and the outflow is powered by the neutrino source.  

The relatively sensitive dependence of the activation condition on the angular momentum of the hole, derived in section \ref{sec:crit-load}, suggests that outflows from slowly rotating black holes may be underpowered.  This may explain the claimed radio loud/quiet dichotomy in AGNs \cite{SSL07}, as discussed in some greater detail at the end of section \ref{sec:discussion}.

\section{\label{sec:eqs}A model for ideal MHD flow with plasma injection}

\begin{figure}
 \includegraphics[width=9.3cm]{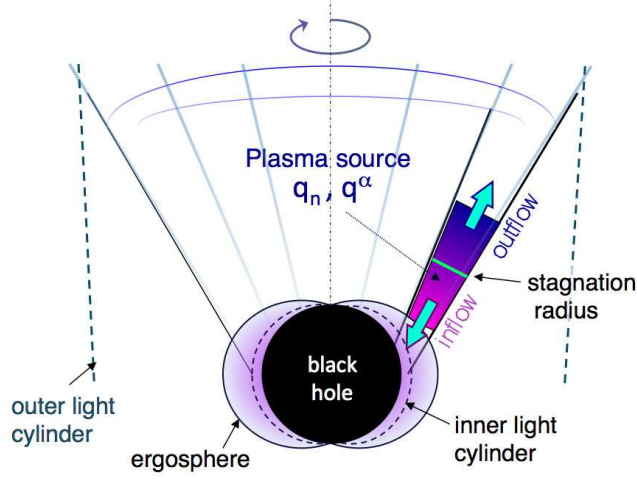}%
 \caption{\label{fig1}  A sketch of the flow structure along a particular streamline:  A double trans-magnetosonic, plasma  flow is launched from a stagnation radius located between the inner and outer light cylinders.  The lost plasma is replenished by an external plasma source, as indicated.   At sufficiently small injection rates, the specific energy of the inflowing plasma is negative, whereas that of the outflowing plasma is positive, implying an outward flow of energy from the horizon to infinity.   This type of flow is powered by the black hole spin energy.  At high injection rates the specific energy is positive everywhere, implying a sign change of the energy flux and the toroidal magnetic field across the stagnation point.  This type of flow is powered entirely by the external plasma source, with some fraction of the injected energy  being absorbed by the black hole and the rest used to accelerate the outflow. }
 \end{figure}

The strong gravitational field of the black hole imposes an inward motion of plasma very near the horizon, regardless of the direction of the energy flux. 
On the other hand, under the conditions suitable for formation of a MHD outflow, the plasma above the outer light cylinder must be flowing outwards.  Consequently, 
the particle flux flowing along magnetic field lines threading the ergosphere must always reverse its direction in the region located between the inner and outer light cylinders \cite{Lev06a}.  Hence, a complete treatment of MHD outflows in Kerr geometry requires proper account of plasma injection in the magnetosphere.   In principle, one can envisage situations in which an outflow cannot be launched in the first place.  For instance,  dumping large amounts of mass at some arbitrary radius above the black hole, e.g., fallback matter from a stellar envelope in collapsars, will disable activation of the BZ process, giving rise to formation of a quasi-steady accretion shock \cite{KB09}.  Such situations are not considered in what follows.  Rather, we focus on cases where the system adjusts to sustain a steady, continuous flow. 
The model constructed below incorporates, in a self-consistent manner, a prescribed plasma source in the flow,  as illustrated in Figure \ref{fig1}.  This source may be associated with mass injection on magnetic field lines, or pair production via annihilation of gamma rays in AGNs and microquasars (e.g., \cite{LR11}) , and neutrinos in GRBs \cite{PW99,zB11}, that emanate from the surrounding accretion disk.  While the later injection processes are well understood and can be accurately modeled, the process of mass injection is only poorly understood.  Mass loading in GRB outflows may conceivably occur via leaking of free neutrons from the hot matter surrounding the jet \cite{LE03}, instabilities at the jet interface, or pick up of baryons from the inner disk.   The last two processes may also be relevant to AGNs and microquasars.  As described below, the MHD equations can be reduced to a system of equations governing the changes in mass, energy and angular momentum fluxes in terms of the corresponding source terms.   The steady double-flow emanating from the stagnation point (see Figure \ref{fig1}) must pass smoothly through the inner and outer fast-and-slow magnetosonic points, the locations of which depend, quite generally, on the energy and momentum deposition profiles.

\subsection{Basic equations}
The stress-energy tensor of a magnetized perfect fluid takes the form,
\begin{equation}
T^{\alpha\beta} = \bar{h}\rho c^2u^{\alpha}u^{\beta} + pg^{\alpha\beta} + \frac{1}{4\pi}\left( F^{\alpha\gamma}F^{\beta}_{\gamma}-\frac{1}{4}g^{\alpha\beta}F^2 \right),
\label{T_M}
\end{equation}
here $u^{\alpha}$ is the four-velocity measured in units of c, $\bar{h}=(\rho c^2+ e_{int} +p) /{\rho c^2}$ the dimensionless specific enthalpy, $\rho$ the baryonic rest-mass density, $p$ the pressure, and $g_{\mu\nu}$ the coefficients of the metric tensor of the Kerr spacetime. In the following we use geometrical units ($c=G=1$), unless otherwise stated, and express the Kerr metric in the regular Boyer-Lindquist coordinates,
\begin{eqnarray}
ds^{2} &\equiv& g_{\mu\nu}dx^{\mu}dx^{\nu} \nonumber \\ &=&-\alpha^{2}dt^{2}+g_{\varphi\varphi}(d\varphi-\omega dt)^2+g_{rr}dr^2+g_{\theta\theta}d\theta^2 
\end{eqnarray}
where the metric coefficients can be expressed as $g_{rr}={{\Sigma}/{\Delta}}$, $g_{\theta\theta}={\Sigma}$, and $g_{\varphi\varphi}\equiv\varpi^2={A}\sin^2\theta/\Sigma$,  in terms of $\Delta=r^{2}+a^{2}-2mr$, $\Sigma=r^{2}+a^{2}\cos^{2}\theta$, and $A=(r^{2}+a^{2})^{2}-a^{2}\Delta \sin^{2}\theta$. The parameters $m$ and $a$ are the mass and specific angular momentum per unit mass of the hole, with $m\geq |a|$.  The coefficients $\alpha=\sqrt{\Sigma\Delta/A}$ and $\omega={2mra}/{A}$ measure, respectively, the time lapse and the frame dragging potential between a zero-angular-momentum observer (ZAMO) and an observer at infinity.  The angular velocity of the black hole is defined as the value of $\omega$ on the horizon, viz.,  $\Omega_H\equiv \omega(r=r_H)=a/(2mr_H)$, here $r_H=m+\sqrt{m^2-a^2}$ is the radius of the horizon, obtained from the condition $\Delta_H=0$.

The dynamics of the flow is governed by the energy-momentum equations:
\begin{equation}
\frac{1}{\sqrt{-g}}(\sqrt{-g}T^{\alpha\beta})_{,\alpha}+
\Gamma^{\beta}_{\ \mu\nu}T^{\mu\nu} = q^\beta, \label{Ttot=q}
\end{equation}
mass conservation:
\begin{equation}
\frac{1}{\sqrt{-g}}\partial_\alpha(\sqrt{-g}\rho u^\alpha)=q_n,\label{continuity-eq}
\end{equation}
and Maxwell's equations:
\begin{eqnarray}
F^{\beta\alpha}_{;\alpha}=\frac{1}{\sqrt{-g}}(\sqrt{-g} F^{\beta\alpha})_{,\alpha}=4\pi j^\beta,\\
F_{\alpha\beta,\gamma}+F_{\beta\gamma, \alpha}+ F_{\gamma\alpha,\beta}=0.\label{Maxwell-homg}
\end{eqnarray}
Here, $q^{\beta}$ denotes the source terms associated with energy-momentum transfer by an external agent, $q_n$ is a particle source, 
and $\Gamma^\beta_{\mu\nu}$ denotes the affine connection.    The magnetic field components measured by a ZAMO are given by
 $B_r=F_{\theta\varphi}/\sqrt{A}\sin\theta$,  $B_\theta=\sqrt{\Delta}F_{\varphi r}/\sqrt{A}\sin\theta$ and  $B_\varphi=\sqrt{\Delta}F_{r\theta}/\Sigma$ \cite{TMP86}.
To simplify the notation we find it useful to define a redshifted poloidal magnetic field: $B_p=(B_r^2+B_\theta^2)^{1/2}/\alpha$.

We consider a stationary and axisymmetric MHD flow in the limit of infinite conductivity, $F_{\alpha\beta} u^\beta=0$. 
In general, the flow is characterized by a stream function $\Psi(r,\theta)$ that defines the geometry of magnetic flux surfaces, and by the following functionals of $\Psi$: 
The angular velocity of magnetic field lines $\Omega(\Psi)$, the  ratio of mass and magnetic fluxes $\eta(\Psi)$, and the energy, angular momentum and entropy  per baryon, denoted by  ${\cal E}(\Psi)$,  ${\cal L}(\Psi)$ and $s(\Psi)$, respectively.  These quantities can be expressed in terms of 
the poloidal velocity, $u_p=\pm(u_r u^r+u_\theta u^\theta)^{1/2}$, where the plus sign applies to outflow lines and the minus sign to inflow lines,  the redshifted poloidal magnetic field $B_p$,  and the azimuthal magnetic field $B_\varphi$ as \cite{Cam86, HMO95,VPL12}:
\begin{eqnarray}
\eta(\Psi)=\frac{\rho u_p}{B_p},\label{eta}\\
\Omega_F(\Psi) = v^\varphi-\frac{v_pB_\varphi}{\varpi B_p},\label{Omega}\\
{\cal E} (\Psi)= -\bar{h}u_t - \frac{\alpha\varpi\Omega_F}{4\pi\eta}B_{\varphi},\label{Enrg}\\
{\cal L} (\Psi)=  \bar{h}u_\varphi -  \frac{\alpha\varpi B_{\varphi}}{4\pi\eta }.\label{Lmom}
\end{eqnarray}
In Equation (\ref{Omega}) $v^\varphi=u^\varphi/u^t$ and $v_p=u_p/\gamma$, with $\gamma=u^t\alpha$ being the Lorentz factor measured by a ZAMO. 
Note that with our sign convention the value of $\eta$ is positive on outflow lines and negative on inflow lines. 
The ideal MHD condition readily implies that $\Omega(\Psi)$ is conserved on magnetic flux surfaces.   
The other quantities are conserved only when $q_n=q^\alpha=0$.  In the general case, the rate of change of $\eta$, ${\cal E}$, ${\cal L}$, and $s$ along streamlines is dictated by Equations  (\ref{Ttot=q})-(\ref{Maxwell-homg}).  
From Ohm's law, $F_{\varphi \mu}u^\mu=0$, and Equation (\ref{continuity-eq}) one obtains
\begin{equation}
u^\alpha\partial_\alpha\eta=\frac{u_p q_n}{B_p}.\label{deriv-eta}
\end{equation}
Likewise, contracting $g_{\beta\gamma}$ with Equation (\ref{Ttot=q}), using the relation $(\sqrt{-g}g^{\alpha\beta})_{,\alpha}+\sqrt{-g}
\Gamma^\beta_{\mu\nu}g^{\mu\nu}=0$, taking the $t$  and $\varphi$ components, and noting that $\Gamma_{\mu t \nu}u^\mu u^\nu=0$ for a stationary flow, yields
\begin{eqnarray}
\frac{1}{\sqrt{-g}}\partial_\alpha(\sqrt{-g}\epsilon^\alpha)=-q_t, \label{e-flux-derv}\\
\frac{1}{\sqrt{-g}}\partial_\alpha(\sqrt{-g}\l^\alpha)=q_\varphi,\label{L-flux-derv}
\end{eqnarray}
where the energy and angular momentum fluxes are given by $\epsilon^\alpha\equiv -T^\alpha_t=\rho u^\alpha{\cal E}$ and  $\l^\alpha\equiv T^\alpha_\varphi=\rho u^\alpha{\cal L}$, respectively.  Finally, the change in the entropy flux, $s^\alpha=(\rho/m_N) u^\alpha s$, where $m_N$ is the nucleon rest mass and $s$ denotes that dimensionless entropy per baryon, is obtained by contracting $u_\beta$ with Equation (\ref{Ttot=q}):
\begin{equation}
\frac{kT}{\sqrt{-g}}\partial_\alpha(\sqrt{-g}s^\alpha)=-u_\alpha q^\alpha.\label{s-flux}
\end{equation}

The normalization condition $u^\alpha u_\alpha=-1$ yields the relation $1+u_p^2=(\alpha u^t)^2-\varpi^2(u^\varphi-\omega u^t)^2$.   By employing Equations (\ref{Enrg}) and (\ref{Lmom})
the latter condition can be written in the form given by Equation (\ref{mot}).    Differentiating the latter equation along a given streamline yields 
\begin{equation}
\left(\ln u_p\right)' = \frac{N}{D}\label{lnupderv}
\end{equation}
where the prime denotes derivative along the streamline $\Psi=$ const, and $N$ and $D$ are given explicitly in the appendix.

\subsection{Flow geometry} 
In a self-consistent treatment, the stream function $\Psi(r,\theta)$ is obtained by solving the trans-field equation.  Such an analysis is beyond the scope of this paper.  To evaluate the conditions required for the activation of the BZ process we invoke, in what follows, a split-monopole configuration.   Such a configuration can be described by a stream function of the form $\Psi(r,\theta)=\Psi_0(1-\cos\theta)$.   With this choice the redshifted poloidal field is given by $B_p = \Psi_0/(2\pi \sqrt{\Sigma\Delta})$.  The poloidal velocity is given by $u_p=\sqrt{\Sigma/\Delta}u^r$, and the convective derivative reduces to $u^\alpha\partial_\alpha = u^r\partial_r=\sqrt{\Delta/\Sigma}u_p\partial_r$.  The energy and angular momentum  fluxes have only a radial component:
\begin{eqnarray}
\epsilon^r=\rho {\cal E} u^r=\frac{\Psi_0}{2\pi \Sigma}\eta{\cal E},\label{eflux-def}\\
\l^r=\rho {\cal L} u^r=\frac{\Psi_0}{2\pi \Sigma}\eta{\cal L}.
\end{eqnarray}

In the next section we show that the sign of the energy flux on the horizon, $\epsilon_H^r$, or equivalently $\eta_H{\cal E}_H$, determines some properties of the flow.

\section{\label{sec:2-types}Two types of flows }

The nature of the flow depends on the rate at which energy (including rest mass energy) is deposited on magnetic field lines.  We identify two distinct types of solutions, that correspond to regimes where the BZ process is switch-on or switch-off.  As we now show, these two types of solutions are characterized by the sign of the specific energy ${\cal E}$ on the horizon.

Let $r_{st}$ denotes the stagnation radius, where $u_p=\eta=0$.    Then, for the double trans-magnetosonic flow considered here  $\eta(r)<0$ at $r<r_{st}$ and $\eta(r)>0$ at $r>r_{st}$.  
Substituting Equation (\ref{eflux-def}) into Equation (\ref{e-flux-derv}) and integrating over $r$ we have
\begin{eqnarray}
\eta(r){\cal E}(r)=\eta_H{\cal E}_H+\frac{2\pi}{\Psi_0}\int_{r_H}^r (-q_t) \Sigma dr^\prime, \quad {\rm at}\quad  r<r_{st},\label{flux-inflow}\\
\eta(r){\cal E}(r)=\eta_\infty{\cal E}_\infty-\frac{2\pi}{\Psi_0}\int_{r}^{r_\infty}(-q_t)\Sigma dr^\prime, \quad {\rm at}\quad  r>r_{st}.\label{flux-outflow}
\end{eqnarray}
The subscripts $H$ and $\infty$ denote the values of quantities on the horizon and at infinity, respectively. The integrals on the right hand side of Equations (\ref{flux-inflow}) and (\ref{flux-outflow}) are associated with energy injection by the external source and, therefore, must be positive. Likewise, ${\cal E}_\infty>0$ always.  Thus, $\eta{\cal E}>0$ at $r>r_{st}$ for both types of flows.    Now, below we show that when $0<\Omega_F<\Omega_H$ and the inertia of the injected matter is sufficiently low, the specific energy on the horizon is negative,  ${\cal E}_H<0$.  In that case $\eta_H{\cal E}_H>0$,  and from Equation  (\ref{eflux-def}) also $\epsilon^r_H>0$, implying that energy  is extracted from the black hole.  From Equation (\ref{flux-inflow}) it is seen that the energy flux at the stagnation radius must be finite, that is $\eta_{st}{\cal E}_{st}>\eta_H{\cal E}_H>0$.   This means that the energy per baryon diverges at $r=r_{st}$; specifically ${\cal E}(r_{st}-\epsilon)\rightarrow -\infty$, and ${\cal E}(r_{st}+\epsilon)\rightarrow +\infty$.  The singularity of the specific energy at $r_{st}$ is a consequence of the fact that the total energy flux there is purely electromagnetic\footnote{This can be directly seen by applying  Equation (\ref{Enrg}) at $r_{st}$ after multiplying by $\eta$, and using $\eta_{st}=0$.}.  From Equation (\ref{mot}) we have
\begin{equation}
{\cal E}_{st}-\Omega_F{ \cal L}_{st}=\bar{h}_{st}\sqrt{k_{ost}},
\label{Es-Ls}
\end{equation}
yielding $\tilde{L}_{st}\equiv {\cal L}_{st}/{\cal E}_{st}=\Omega_F^{-1}$.
The azimuthal magnetic field at $r_{st}$ can be readily obtained from Equation (\ref{Enrg}):
\begin{eqnarray}
B_{\varphi}(r_{st})=-\frac{4\pi\eta_{st }\,{\cal E}_{st}}{\alpha_{st}\varpi_{st}\Omega_F}, \label{bphi-stg}
\end{eqnarray}
and it is seen that $B_\varphi$ maintains its sign across the stagnation zone.   The above considerations indicate that in this regime the dynamics of the flow is governed by the black hole rotation.  In the force-free limit, in which the inertia of injected matter is negligible, that is, $2\pi\int_{r_H}^{r_\infty}(-q_t)\Sigma dr/(\Psi_0\eta_H{\cal E}_H)\rightarrow 0$, Equations (\ref{flux-inflow}) and (\ref{flux-outflow}) yield $\eta_\infty{\cal E}_\infty=\eta_H{\cal E}_H$, confirming that the spin down power of the black hole is the sole energy source of the  outflow.  Note that the structure of this type of flows is fundamentally different than that of an ideal MHD outflow from a stellar surface (see, e.g., \cite{Mich69,Mstl12, HenRay71}), as there is a region is space where energy is flowing against the plasma stream.   This strange behavior is a unique feature of frame dragging, that allows the existence of two light surfaces; a conventional one located well outside the ergosphere, and an inner one located inside the ergosphere where $g_{tt}>0$ (see appendix \ref{sec:appA} for further details).   As explained above, within the inner light surface particles must travel radially inward along negative energy trajectories. 

As shown below, when loading of magnetic field lines by the external source exceeds a critical value, the specific energy on the horizon becomes positive, ${\cal E}_H>0$.  Then, $\eta_H{\cal E}_H<0$, meaning that the black hole is fed by the external source.   Since $\eta_\infty{\cal E}_\infty>0$, it is evident that the energy flux changes sign in the injection zone, and so must vanish at $r_{st}$; that is, $\eta_s{\cal E}_{st}=0$.   Consequently, the specific energy is finite and continuous at $r_{st}$, unlike the behavior of the previous flow type.  Equation (\ref{bphi}) yields $B_\varphi(r_{st})=0$, implying that $B_{\varphi}$ must also change sign across the stagnation radius.    As seen from Equations (\ref{flux-inflow}) and (\ref{flux-outflow}), $|\eta_\infty{\cal E}_\infty|+|\eta_H{\cal E}_H|=2\pi\Psi_0^{-1}\int_{r_H}^{r_\infty}(-q_t)\Sigma dr$, indicating that the flow is powered by the energy deposited on magnetic field lines alone.  Thus, this type of flow is driven by the external source rather than by the spin energy of the black hole.  The angular velocity $\Omega_F$ is presumably fixed by the rotation of injected matter, as suggested by the fact that $B_\varphi=0$ and $\Omega_F=v^\varphi$ at $r_{st}$ (see Equation (\ref{Omega})).  
The properties of the outflow emanating from the stagnation radius are similar in some respects to those of outflows ejected from a stellar surface or an accretion disk.  Sufficiently far out they may be well described by the Michel's solution \cite{Mich69, HenRay71} if they are sufficiently magnetized. 
A particular example of such a flow with $a=\Omega_F=0$ and a realistic energy deposition profile is outlined in \cite{LG13}.

\section{\label{sec:crit-load}A critical load}

To simplify the analysis we suppose that the injection zone is infinitely thin, that is $q^\alpha(r) \propto \delta(r-r_{st})$, and likewise $q_n$.    Since we are merely interested here in evaluating the dependence of the energy flux at the horizon, $\epsilon^r_H$, on the load, it is sufficient to consider the inflow section in the region  $r_{H}< r< r_{st}$.    For the injection model adopted here Equations (\ref{deriv-eta})-(\ref{s-flux}) imply that $\eta$, ${\cal E}$, ${\cal L}$ and $s$ are conserved on magnetic surfaces in the region  $r_{H}< r< r_{st}$.   The structure of the flow is then obtained upon integration of Equation (\ref{lnupderv}).  To elucidate key features, we present results obtained in two extreme limits: a cold flow and a relativistically hot flow.   

\subsection{Cold flow}
We consider first a cold adiabatic flow. We set $\bar{h}=1$ and note that in the absence of plasma injection ($q^\alpha=q_n=0$) the location of the slow magnetosonic point of a cold flow, $r_{sm}$, coincides with the stagnation radius, that is,  $u_p(r_{sm})=u_{sm}=0$.
As argued by \cite{TNTT90}, the requirement that $u_p^\prime$ remains finite at the slow point, where $D=0$, implies $k_0^\prime=0$ there.  This can be readily verified by taking the limit $a_s^2\rightarrow 0$,  $u_p\rightarrow u_{sm}$ in Eqs. (\ref{lnupderv_start})-(\ref{lnupderv_end}).  For the split monopole geometry adapted here this condition reads:
\begin{equation}
\frac{d}{dr}\left[ \alpha^2 - \varpi^2\left(\Omega_F-\omega\right)^2\right]=0.
\label{slow-rad-cold}
\end{equation}
The solution of the latter equation gives  the slow magnetosonic radius on every streamline, $r_{sm}(\theta)$.     In general, the stagnation radius $r_{st}$ does not coincide with $r_{sm}$, meaning that the slow point is located inside the injection zone, where the above analysis breaks down.   Moreover, the exact shape of 
magnetic surfaces should depend on the details of the plasma injection process (although we anticipate small deviations from the split monopole configuration adopted here in the regime of small inertia).    In the following, we ignore these complications and restrict our analysis to radial inflows.  We note that 
for every choice of $\eta$, $\Omega_F$ and $\theta$ there exists a unique solution outside the injection zone, in the region $r_{H}< r< r_{st}$, that passes smoothly through the fast magnetosonic point.  Each such solution can be extrapolated to the radius $r_{sm}>r_{st}$ where the boundary condition $u_p=0$ can be used.   This procedure is not mandatory, and has been used for convenience.  The value of $\eta$  at $r=r_{st}-\epsilon$ depends on the particle source $q_n$;  for $q_n(r)=q_{n0}\delta(r-r_{st})$ we obtain from Equation (\ref{deriv-eta}) 
\begin{equation}
\eta=\frac{q_{n0}}{B_{pst}}\sqrt{\Sigma_{st}/\Delta_{st}}.
\label{eta-from-q_n}
\end{equation} 
The implicit assumption underline our analysis is that the acceleration of the flow within the injection zone is consistent with the boundary conditions at $r_{st}$. For our simple injection model this condition can be fulfilled  for appropriate choice of the source terms $q^\alpha$.  Self-consistent calculations of double trans-magnetosonic flows with realistic injection profiles will be presented in a follow-up paper. 

\begin{figure}
 \includegraphics[width=9.3cm]{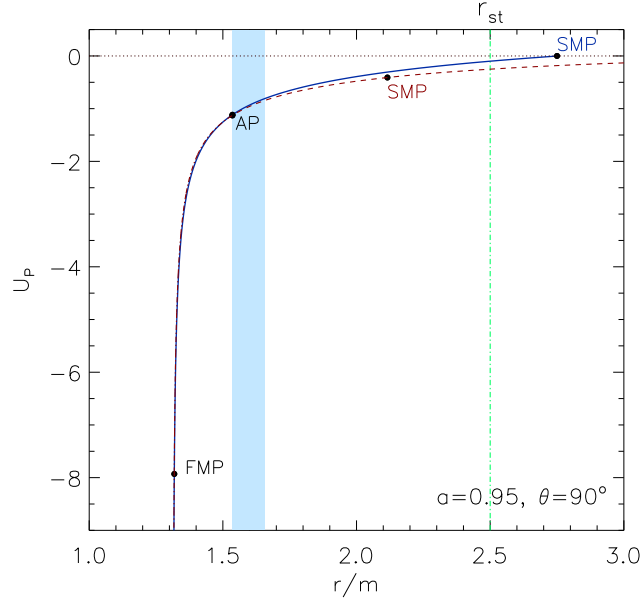}%
 \caption{\label{fig2} Radial profiles of the poloidal velocity, $u_p$, of a cold (solid line) and a relativistically hot (dashed line) negative energy  flows with a radial magnetic field.   The slow-magnetosonic, Alfv\'en, and fast-magnetosonic points are indicated by SMP, AP, and FMP, respectively.    The fast-magnetosonic point of the hot flow is located at $(r/m=1.315, u_p= -11.53)$, and is not shown.  The stagnation radius, $r_{st}$, is marked by the vertical dotted-dashed line.  The blue shaded region delineates the permitted range of Alfv\'en radii of all negative energy solutions.}
 \end{figure}

We seek solutions that describe an inflow of plasma into the black hole ($u_p\le 0$).  For a given choice of the black hole parameters $a$ and $m$, magnetic flux $\Psi_0$, and  angular velocity $\Omega_F$, this family of solutions is characterized by $\eta$ and  $\theta$. 
For a given choice of $\eta$, a  solution is obtained by integrating Equation (\ref{lnupderv}) along a streamline defined by $\theta=\theta_0$.  The integration starts at $r_{sm}(\theta_0)$, which we compute first using Equation (\ref{slow-rad-cold}), and is repeated  iteratively by changing the value of ${\cal E}$  until a smooth transition across the fast magnetosonic point is achieved.  The value of $\tilde{L}$ is computed,  in every run, from Equation (\ref{Es-Ls}). 
A typical negative energy inflow solution, computed using $\eta=0.023\,\textrm{g cm}^{-2} \textrm{s}^{-1} \textrm{G}^{-1}$, 
$a/m=0.95$, $\Omega_F=\Omega_H/2$, $\theta_0=90^{\circ}$,
is displayed in figure \ref{fig2} (solid line). It starts from the slow magnetosonic radius ($r_{sm}=2.75\,m$), denoted SMP in the figure, and passes through 
the Alfv\'en and the fast magnetosonic points, denoted AP and FMP, respectively. 

\begin{figure*}
 \includegraphics[width=15.3cm]{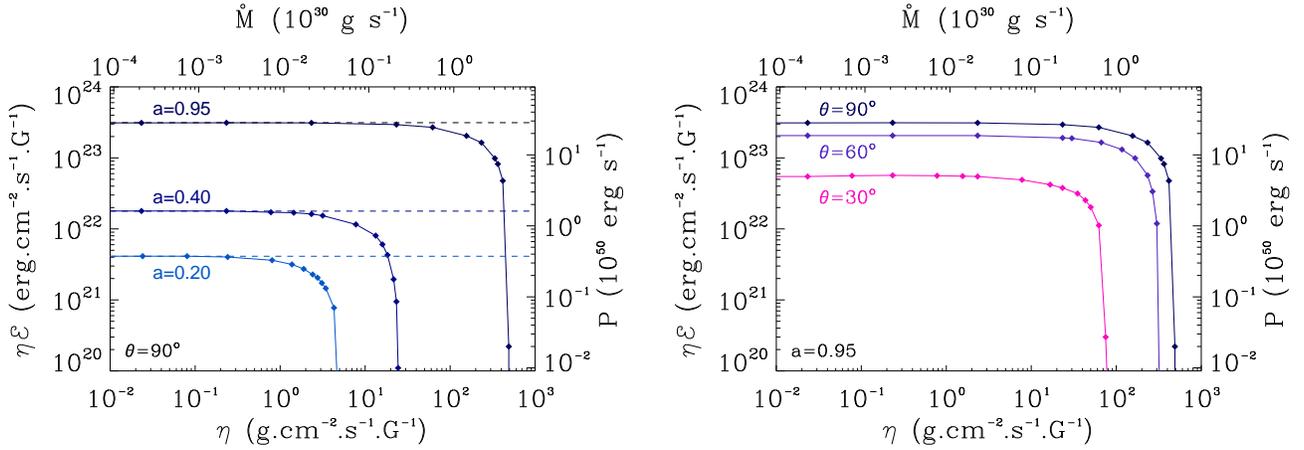}%
 \caption{\label{fig3} Total energy flux $\eta {\cal E}$ vs. mass-to-magnetic flux ratio $\eta$, in the regime where energy extraction is switched on.  Each point corresponds to a cold, negative energy solution, like the example shown in figure 2. The values of the mass flux $\dot{\cal M}(\theta)$ and power $P(\theta)$, defined in Equations (\ref{mdot2}) and (\ref{P_j(Q)}), are given in the top and right axis, respectively, for $\Psi_0\approx9\times10^{27}$ G cm$^2$.  The horizontal dashed lines shown in the left panel correspond to the BZ power of a force-free flow, given explicitly in Equation (\ref{eflux-ff}).}
 \end{figure*}

Figure \ref{fig3} delineates the dependence of $\eta{\cal E}$ on the mass-to-magnetic flux ratio $\eta$ in the regime where energy extraction is swiched on (${\cal E}<0$), for different values of $a$ and $\theta$.  For convenience, we give also the values of the angular distribution of the mass flow rate and extracted power, defined here as 
\begin{equation}
\dot{\cal M}(\theta)=2\pi\Sigma \rho u^r=\eta\Psi_0,
\label{mdot2}
\end{equation}
and 
\begin{equation}
P(\theta)=2\pi\Sigma \epsilon^r=\dot{\cal M}{\cal E},
\label{P_j(Q)}
\end{equation}
respectively.  The horizontal dashed lines mark the analytic result derived by BZ in the force-free limit for $\Omega_F=\Omega_H/2$:
\begin{equation}
P_{FF}(a,\theta)=\frac{c}{128\pi^2}\left(\frac{a}{m}\right)^2\frac{(r_H^2+a^2)\sin^2\theta}{r_H^2(r_H^2+a^2\cos^2\theta)}\Psi_0^2\,,
\label{eflux-ff}
\end{equation}
and it is seen that the extracted power converges to this limit at sufficiently small loads, but is strongly suppressed as the load approaches the critical value $\dot{\cal M}_c= P_{FF}/c^2$ (or $\eta_c=P_{FF}/\Psi_0$), and eventually switched off.

In order to compare our result with the test simulations of \cite{KB09}, we employ Equations (\ref{mdot2}) and (\ref{eflux-ff}) to write
\begin{equation}
\frac{P_{FF}}{\dot{\cal M}c^2}=\frac{(r_H^2+a^2)\sin^2\theta}{8(r_H^2+a^2\cos^2\theta)}\kappa^2,
\end{equation}
where $\kappa$ is the parameter defined in Equation (6) of \cite{KB09}.   From figure \ref{fig3} we find the activation condition to be $P_{FF}/\dot{\cal M}c^2>0.5$ on the equatorial plane for $a=0.95$, which corresponds to $\kappa>2$, in a good agreement with \cite{KB09}. 

\subsection{Hot flow}

Next, we generalize the above analysis to a hot flow,  $\bar{h}>1$.   We assume that the pressure $p$ is dominated by radiation, and set  $w=\rho\bar{h}=\rho +4p$.
Unlike in the case of a cold inflow, the slow magnetosonic point of a hot inflow is located below the stagnation radius, at $r_{sm}<r_{st}$, and is unknown a priori.    Thus, our strategy is to start the integration  of Equation (\ref{lnupderv}) at some radius below $r_{sm}$, and seek solutions that pass smoothly through both, the slow-and-fast magnetosonic points.    A typical negative energy, hot inflow solution is delineated by the dashed line in figure 2.
The family of solutions thereby computed is characterized by the parameter  $(wu_p/B_p)_{sm}$, that denotes the enthalpy flux per unit magnetic flux at the slow magnetosonic point, and which reduces to $\eta$ at zero temperature.   In the spirit of Equation (\ref{mdot2}) we define the quantity
\begin{equation}
\dot{w}_{sm}(\theta)=\Psi_0(wu_p/B_p)_{sm}=(2\pi\Sigma w u^r)_{sm},
\end{equation}
which approaches $\dot{\cal M}(\theta)$ in the limit $\bar{h}\rightarrow1$.  As shown in figure \ref{fig3}, the effect of the load on the extracted power can be quantified in terms of this parameter. 

The specific entropy of a relativistically hot gas is given approximately by $s=w/(nkT)$.  Substituting the latter relation into Equation (\ref{s-flux}), and adopting for simplicity $q^\alpha=\dot{Q}_0\delta(r-r_{st})[1,0,0,0]$,  we obtain $(wu^r)_{st}\simeq \gamma_{st}\dot{Q}_0$.  Since the enthalpy flow rate, $2\pi\Sigma w u^r$, barely changes along streamlines, and since $\gamma_{st}\simeq1$, we have approximately:
\begin{equation}
\dot{w}_{sm}\simeq2\pi\Sigma_{st}\dot{Q}_{0}\simeq P_{inj}(\theta),
\label{P_inj}
\end{equation} 
where $P_{inj}(\theta)=d\dot{E}_{ext}/d(\cos\theta)$ denotes the angular distribution of the power deposited in the magnetosphere by the external energy source.  

\begin{figure}
 \includegraphics[width=9.2cm]{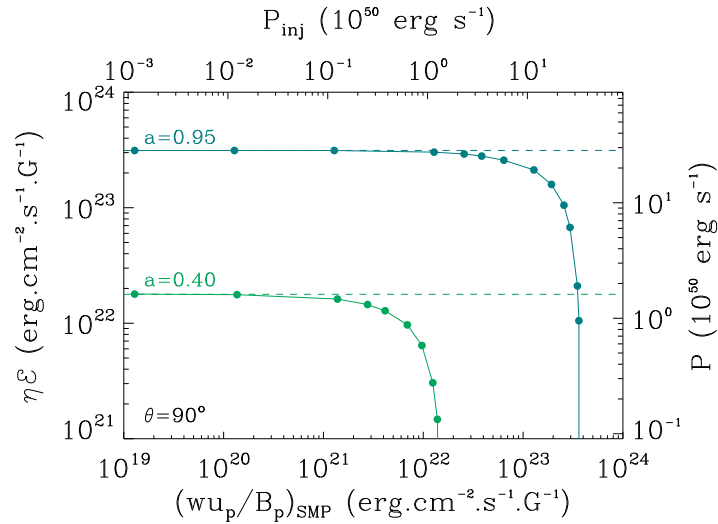}%
 \caption{\label{fig4} Dependence of $\eta {\cal E}$ on the enthalpy flux per unit magnetic flux at the slow magnetosonic point, $(wu_p/B_p)_{sm}$, computed
for a family of relativistically hot, negative energy solutions.  The upper axis gives the injected power, Equation (\ref{P_inj}), and the right axis the extracted power defined in Equation (\ref{P_j(Q)}).}
 \end{figure}

Figure \ref{fig4} exhibits the dependance of the outgoing energy flux $\eta{\cal E}$ on $(wu_p/B_p)_{sm}$ for $\theta=\pi/2$ and different values of $a$.    As seen from the figure, the critical condition for activation of the BZ process is $\dot{w}_{sm}<P_{FF}$ or, using Equation (\ref{P_inj}), $P_{inj}(\theta)<P_{FF}(a,\theta)$ .   This condition generalizes the cold flow result, for which $P_{inj}=\dot{\cal M}c^2$, to a flow with arbitrary temperature.

\section{\label{sec:discussion}Discussion} 

The above results indicate that the rotational energy of a Kerr black hole can be magnetically extracted provided that the rate at which energy is deposited on magnetic field lines by the plasma source does not exceed the BZ power of a force-free flow,  given explicitly in Equation (\ref{eflux-ff}).     In the case of a cold plasma, this condition reduces to a limit on the mass flux flowing into the black hole along a magnetic surface.  When expressed in terms of the angular distribution of the mass flow rate, $\dot{\cal M}(\theta)=d\dot{M}/d(\cos\theta)$, this critical condition reads: 
\begin{equation}
\dot{{\cal M}(\theta)}< 10^{-4}\left(\frac{M_{BH}}{3M_{\odot}}\right)^{-2}\left(\frac{\Psi_0}{10^{27}{\rm G\,cm^2}}\right)^{2}g(a,\theta)\quad {\rm M_{\odot}\, s^{-1}},
\label{activation-cond}
\end{equation}
where $g(a,\theta)=a^2(r_H^2+a^2)\sin^2\theta/[r_H^2(r_H^2+a^2\cos^2\theta)]$.  

A rough estimate of the maximum magnetic flux that can be accumulated near the horizon of the black hole in a GRB engine can be obtained using the disk model of \cite{PW99}, and assuming equipartition of gas and magnetic pressure:
\begin{eqnarray}
\Psi_{max}\simeq 10^{29} \left(\frac{\alpha_{viss}}{0.1}\right)^{-0.55}\left(\frac{M_{BH}}{3 M_{\odot}}\right)^{1.05}&&\left(\frac{\dot{M}_{acc}}{M_{\odot}\ s^{-1}}\right)^{0.5} \,{\rm G\ cm^2},\label{Bflux-from-disk}
\end{eqnarray}
here $\alpha_{viss}$ and $\dot{M}_{acc}$ denote the viscosity parameter and accretion rate of the neutrino-cooled accretion flow, respectively.  Equation (\ref{Bflux-from-disk})  largely overestimates the actual  value of the flux that is likely to be accumulated.  Firstly, more realistic disk models \cite{CB07} yield a smaller pressure in the inner disk regions and, hence, smaller $\Psi_{max}$, by about an order of magnitude.  Secondly, only some fraction of this maximum value is accumulated in practice.    We anticipate $\Psi_0 \simlt 10^{28}$ G cm$^2$ even at accretion rates approaching $\sim 1$ M$_{\odot}$ s$^{-1}$.  This implies that along field lines that extract energy from the black hole, mass inflow must be strongly suppressed.  Suppression of the baryon load is expected in the polar region by virtue of the angular momentum barrier.  But even then, the requirements for energy extraction and formation of a relativistic outflow impose stringent constraints on the progenitors, as discussed in \cite{KB09}.     

Another plasma source in GRB jets is annihilation of MeV neutrinos that emanate form the hyper-accretion disk surrounding the black hole.   The plasma thereby deposited is relativistically hot, and so a polar outflow will be driven either by the black hole or by the pressure of the injected plasma, provided that the central region is baryon poor, as explained above.  The type of the outflow will be determined by the energy load of magnetic field lines, as explained in section \ref{sec:2-types}.  
Detailed calculations that exploit an advanced disk model \cite{zB11} yield a net energy deposition rate of $\dot{E}_{\nu\bar{\nu}}\simeq 10^{52} \dot{m}_{acc}^{9/4}\left(\frac{M_{BH}}{3 M_{\odot}}\right)^{-3/2}x_{mso}^{-4.8}$ erg s$^{-1}$, for accretion rates (henceforth measured in units of $M_{\odot}$ s$^{-1}$) in the range $\dot{m}_{ign}<\dot{m}_{acc}<\dot{m}_{trap}$, where $x_{mso}$ is the radius of marginally stable orbit in units of $m$. Assuming for simplicity a uniform angular distribution, viz., $P_{inj}(\theta)=\dot{E}_{\nu\bar{\nu}}/2$, we derive an approximate condition for activation of the BZ process:
\begin{equation}
\dot{m}_{acc}<0.1\left(\frac{M_{BH}}{3M_\odot}\right)^{-2/9}\left(\frac{\Psi_0}{10^{27}{\rm G\ cm^2}}\right)^{8/9}f(a,\theta).\label{active-BZ}
\end{equation}
The  function $f(a,\theta)$ satisfies $f(0,\theta)=0$, but otherwise depends weakly on $a$.  For $\theta=\pi/2$ it varies between 1 and $1.2$ in the range $0.95\ge a\ge 0.2$.   When condition (\ref{active-BZ}) is satisfied, the outflow is powered by the spinning black hole.  When this condition is violated, the flow is driven by the pressure of the e$^\pm$ pairs produced in the magnetosphere.   In the latter case, a fraction of the injected power will emerge at infinity in the form of a relativistic outflow, and the rest will get absorbed by the black hole.   A particular example of such a double-transonic flow is exhibited in \cite{LG13}. 

The relatively sensitive dependence of the switch-on condition on black hole spin (see figure \ref{fig3}), suggests that slowly rotating black holes in AGNs 
(and perhaps also in X-ray binaries) are either quite, or have under-powered outflows.     For instance, if the black hole is surrounded by a thick disk,
then it could be that the inclination angles of magnetic field lines that have a sufficiently low mass load to allow energy extraction depend on the angular momentum of the hole $a$ via
the activation condition (\ref{activation-cond}).  This may result in a steeper dependence of the jet power on $a$ than the usual scaling obtained in the force-free limit,
and may explain the claimed radio loud/quiet dichotomy \cite{SSL07}.  A different, though perhaps related, explanation for this dichotomy has been offered by \cite{TNM10}.

\begin{widetext}
\appendix*
\section{\label{sec:appA}Derivation of the equation of motion of the MHD flow}

Defining $M$ as the poloidal Alfv\'enic Mach number through $M^2\equiv{4\pi \bar{h} {\eta}^2 c^2}/{\rho}={u_p^2}/{u_A^2}$,
where $u_A^2=B_p^2/(4\pi \bar{h} \rho c^2)$, and given the following expressions:
\begin{eqnarray}
k_0 &=& \alpha^2 - \varpi^2\left(\Omega_F-\omega\right)^2,\\
k_2 &=& ({\cal E}-\Omega_F {\cal L})^2,\\
k_4 &=& \frac{{\cal L}^2}{\varpi^2}-\frac{\left({\cal E}-\omega{\cal L}\right)^2}{\alpha^2}\,,
\end{eqnarray}
we can use the constants of motion (\ref{eta})-(\ref{Lmom}) and the normalization condition of the 4-velocity, $u^\alpha u_\alpha=-1$,  to obtain algebraic relations for the 4-velocity components:
\begin{eqnarray}
&&u^t=\frac{\alpha^2 ({\cal E}-\Omega_F {\cal L})-M^2({\cal E}-\omega{\cal L})}{\alpha^2\bar{h}\left(k_0-M^2\right)}\,,\label{ut} \\
&&u^\varphi= \frac{\alpha^2 \Omega_F({\cal E}-\Omega_F {\cal L})-M^2\omega({\cal E}-\omega{\cal L})-{M^2L\alpha^2}{\varpi^{-2}}}{\alpha^2\bar{h}\left(k_0-M^2\right)}\,,\\\label{uphi}
&&u_p^2+1=\frac{k_2\left(k_0  - 2  M^2\right) -k_4 M^4}{\bar{h}^2(k_0-M^2)^2}\label{mot}\,.
\end{eqnarray}

We also express the toroidal component of the magnetic field as
\begin{eqnarray}
B_{\varphi}=-\frac{4\pi\eta}{\alpha\varpi} \;\frac{\alpha_{}^{2}{\cal L}-\varpi^{2}(\Omega_F-\omega)({\cal E}-{\cal L}\omega)}{k_0-M^{2}}\label{bphi}\,.
\end{eqnarray}

Equation (\ref{mot}) is the wind equation for the poloidal velocity \cite{Cam86,TNTT90}.  Upon differentiating this equation along a given streamline, $\Psi=$const, one obtains the equation of motion
\begin{equation}
\left(\ln u_p\right)' = \frac{N}{D}\,,\label{lnupderv_start}
\end{equation}
with
\begin{eqnarray}
N&=& \zeta_1\left(\ln B_p\right)'+\zeta_2\left(\ln \alpha\right)'+\zeta_3\left(\ln \varpi\right)'+\zeta_4\left(\ln {\cal E}\right)'+\zeta_5\left(\ln s\right)'+\zeta_6\left(\ln \omega\right)'\,,\\
D&=&\left(k_0-M^2\right)^2\left[\left({u}_p^2-c_s^2\right)\left(k_0-M^2\right)+ \frac{M^4}{\bar{h}^2}\frac{\left(k_0 k_4+k_2\right)}{\left(k_0-M^2\right)^2} \right].
\label{Dapp}
\end{eqnarray}
Here $c_s^2$ is the sound 4-velocity defined by $c_s^2=a_s^2/{(1-a_s^2)}$, with $a_s^2$ is  given by Eq. (25) in \cite{Lev06b}, and

\begin{eqnarray}
\zeta_1 &=& -\left(k_0-M^2\right)^2\left[\left(1+u_p^2\right)\left(k_0-M^2\right)c_s^2-M^2\frac{B_\varphi^2}{4\pi\bar{h}\rho}\right]\,,\\
\zeta_2 &=&\frac{1}{\bar{h}^2\left(1-a_s^2\right)}\left\{\frac{M^6\left({\cal E}-\omega{\cal L}\right)^2}{\alpha^2}- \left[ \left({\cal E}-\omega{\cal L}\right)^2\left(3-\frac{\varpi^2\delta\Omega^2}{\alpha^2}\right)-\frac{2\alpha^2{\cal L}^2}{\varpi^2}\right]M^4 + \alpha^2k_2\left(3M^2-k_0\right) \right\}\,,\nonumber\\\\
\zeta_3 &=& \frac{1}{\bar{h}^2\left(1-a_s^2\right)}\left\{-\frac{M^6{\cal L}^2}{\varpi^2}- \left[ 3{\cal L}^2 \delta\Omega^2-\frac{\alpha^2{\cal L}^2}{\varpi^2}-\frac{2\varpi^2}{\alpha^2}\delta\Omega^2\left({\cal E}-\omega{\cal L}\right)^2\right]M^4 -\varpi^2\delta\Omega^2k_2\left(3M^2-k_0\right) \right\}\,,\nonumber\\\\
\zeta_4 &=& \frac{1}{\bar{h}^2\left(1-a_s^2\right)}\left(k_0-M^2\right)\left[\left(k_0-2M^2\right)\left( {\cal E}-\Omega_F{\cal L}\right){\cal E}+\frac{M^4{\cal E}}{\alpha^2} \left({\cal E}-\omega{\cal L}\right)\right]\,,\\
\zeta_5 &=& \frac{s \,c_s^2 \left(5+8\sigma\right)}{\bar{h}^2\left(5+10\sigma+2\sigma^2\right)}\left[-k_4M^6-k_2\left(k_0^2-3k_0M^2+3M^4\right)\right]\,,\\
\zeta_6 &=& -\frac{1}{\bar{h}^2\left(1-a_s^2\right)}\left[M^4\left(k_0-M^2\right)\left({\cal E}-\omega{\cal L}\right)\frac{{\cal L}\omega}{\alpha^2}+\varpi^2\omega\delta\Omega\left(k_0k_2-3k_2M^2-2k_4M^4\right)\right]\,,\label{lnupderv_end}
\end{eqnarray}
generalize the coefficients $\zeta_{i={1,6}}$ derived in \cite{Lev06b} in the Schwarzschild geometry, where for short we denote $\delta\Omega\equiv\Omega_F-\omega$.

\subsection{\label{sec:App-CS}Critical surfaces}

The requirement of a smooth transition between the sub- and-super-alfv\'enic regimes, yields the following regularity condition at the Alfv\'en surface, where the denominator of Eqs. (\ref{ut})-(\ref{bphi}) vanishes:
\begin{eqnarray}
M^2_A = \alpha_A^2 - \varpi_A^2\left(\Omega_F-\omega_A\right)^2\label{AP1},\\
\varpi^2_A = \frac{\alpha_A^2\tilde{L}}{\left(\Omega_F-\omega_A\right)\left(1-\omega_A\tilde{L}\right)},\label{AP2}
\end{eqnarray}
where $\tilde{L}={\cal L}/{\cal E}$.  Equation (\ref{AP1}) has two roots that define the inner and outer Alfv\'en surfaces.  Those surfaces approach the light surfaces in the limit of zero inertia,  at which $M_A^2\rightarrow0$.  The outer light surface is located well outside the ergosphere, where gravity is weak and to a good approximation $\alpha\simeq1$, $\omega=0$.  To lowest order it coincides  with the conventional light cylinder, $\varpi_c\simeq\Omega_F^{-1}$, as originally derived in Ref \cite{Mich69} for pulsar winds.   Recalling that $g_{tt}=-\alpha^2+\varpi^2\omega^2$, and using Equation (\ref{AP1}) with $M_A=0$, gives $\varpi^2\omega^2>g_{tt}=\varpi^2[\omega^2-(\Omega_F-\omega)^2]>0$ at the inner light surface.  Hence, it must be located inside the ergosphere, but above the horizon since $\Omega_F<\omega_H$.  \\

There are two additional critical surfaces, the fast-and-slow magnetosonic surfaces.   Those can be most conveniently identified by expressing the denominator 
$D$ (Equation \ref{Dapp}) in the form,
\begin{eqnarray}
D&=&-\left(k_0-M^2\right)^2\left({u}_p^2-{u}_{sm}^2\right)\left({u}_p^2-{u}_{fm}^2\right)(u_A^2)^{-1},
\end{eqnarray}
in terms of the slow and fast magnetosonic speeds,
\begin{eqnarray}
u^2_{sm}= K-\sqrt{K^2-c_s^2u_A^2k_0}\,,\\
u^2_{fm}= K+\sqrt{K^2-c_s^2u_A^2k_0}\,,\\
\end{eqnarray}
where
\begin{equation}
K=\frac{1}{2}\left[k_0u_A^2+c_s^2+\frac{B_\varphi^2}{4\pi\bar{h}\rho}\right].
\end{equation}

\end{widetext}

\bibliography{basename of .bib file}

\end{document}